\documentclass{ptephy_v1}
\preprintnumber{2108.11566} 
\usepackage{graphicx}
\usepackage{bm}
\usepackage{multirow}
\newcommand{\diff}{{\mathrm d}}

\begin{document}

\title{Non-monotonic behavior of the Binder Parameter in the discrete spin systems}

\author[1,2]{Hiroshi Watanabe}

\affil{Department of Applied Physics and Physico-Informatics, Keio University, Yokohama 223-8522, Japan \email{hwatanabe@appi.keio.ac.jp}}

\author[2]{Yuichi Motoyama}
\affil{The Institute for Solid State Physics, The University of Tokyo, Kashiwanoha 5-1-5, Kashiwa, Chiba 277-8581, Japan}

\author[2]{Satoshi Morita}
\author[2]{Naoki Kawashima}

\date{\today}


\begin{abstract}
    We study a non-monotonic behavior of the Binder parameter, which appears in the discrete spin systems. We show that the Binder parameters of the Potts model are non-monotonic for $q=3$ and $4$, while they are monotonic for the Ising case ($q=2$). Using the Fortuin-Kasteleyn graph representation, we find that the improved estimator of the Binder parameter consists of two terms with values only in high- and low-temperature regions. The non-monotonic behavior is found to originate from the low-temperature term. With the appropriately defined order parameter, we can reduce the influence of the low-temperature term, and as a result, the non-monotonic behavior can also be reduced. We propose new definitions of the order parameter, which reduces or eliminates the non-monotonic behavior of the Binder parameter in a system for which the improved estimator of the Binder parameter is unknown.
\end{abstract}

\maketitle

\section{Introduction}

The Binder parameter, \textit{a.k.a} the forth-order cumulant, is the useful tool for analyzing the critical phenomena~\cite{Binder1981,Binder1985}. Since the scaling dimension of the Binder parameter is zero, this value does not depend on the system size at the criticality. Therefore, the Binder parameters with different system sizes cross at the criticality which allows us to identify the critical point. Not only the spin systems but it can also be applied to particle systems~\cite{Watanabe2012}. Recently, a method for calculating Binder parameters using the tensor network was also proposed~\cite{Morita2019}.
While there are dimensionless variables, such as the ratio of the correlation length to the linear system size, the Binder parameter has been widely used since it is easy to use and exhibits better convergence. However, the Binder parameter sometimes indicates peculiar behavior compared to other dimensionless parameters. For example, the Binder parameters are more affected by finite size effects than $\xi/L$~\cite{Hasenbusch2008, Tomita2002}, where $\xi$ is the correlation length and $L$ is the system size. The correction to scaling manifests itself most prominently as the non-monotonic size-dependence, a hump. The hump appears near the criticality in some systems, such as the Potts model (see Fig.~\ref{fig:binder}). The height of the hump often exhibits a significant system-size dependence, which makes the scaling analysis difficult. While the hump does not appear in the Ising model, the Binder parameters of the $q$-state Potts models with $q>2$ exhibit humps. The humps appear not only in the Potts models but in the frustrated system~\cite{Sandvik2012,Kalz2012}. Humps also appear in quantum systems~\cite{Harada2013, Suzuki2015}. The existence of the hump of the Binder parameter is known among researchers. For example, phenomenological arguments are given that the Binder parameter exhibits a hump when the system involves the first-order transition~\cite{Vollmayr1993}. Jin \textit{et al.} utilized the system-size dependence of the hump to identify the Potts point for the $J_1$--$J_2$ frustrated Ising model~\cite{Sandvik2012}. Kalz and Honecker adopted the Ising-like order parameter, which reduces the effects of the hump~\cite{Kalz2012}. Recently, Patil and Sandvik proposed a new scenario for the origin of the hump of the Binder parameter in the system involving a pseudo-first-order transition~\cite{Patil2020}. They suggested that the nonmonotonic behavior of the Binder parameter originated from the Ising-like fluctuations of the order parameter near the criticality, where the symmetry of the ordered phase is not $Z_2$. Despite these studies, there are still many unanswered questions about the humps of the Binder parameter. For example, it is not known why the effect of the hump is reduced by adopting the Ising-like order parameter or whether the hump can be eliminated.

The purpose of this paper is to reveal the origin of the non-monotonic behavior of the Binder parameter and to propose 
a method for eliminating it. We investigate how different definitions of the Binder ratio affect the results of the finite-size scaling. This paper is organized as follows. In the next section, we describe the properties of humps of the Binder parameter, especially the origin of the hump of the Binder parameter. In Sec.~\ref{sec:elimination}, we discuss eliminating or alleviating the humps with appropriately defined order parameters. In Sec.~\ref{sec:frustratedising}, we demonstrate that an appropriately defined order parameter can eliminate the effect of the hump even in systems where the improved estimator of the Binder parameter is unknown. Section~\ref{sec:summary} is devoted to discussion and perspective.

\section{Hump of Binder Parameter in Potts Models} \label{sec:hump}

\begin{figure}[htbp]
    \includegraphics[width=8cm]{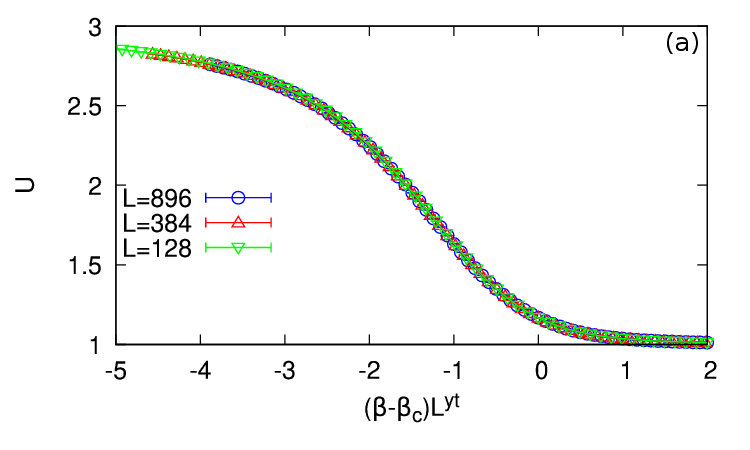}
    \includegraphics[width=8cm]{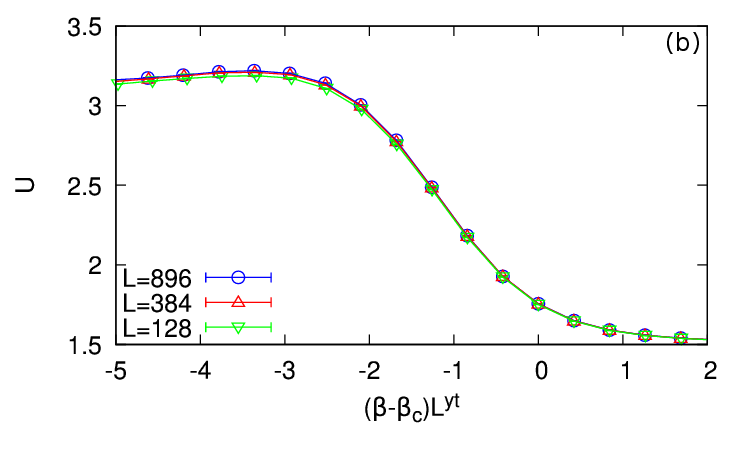}
    \includegraphics[width=8cm]{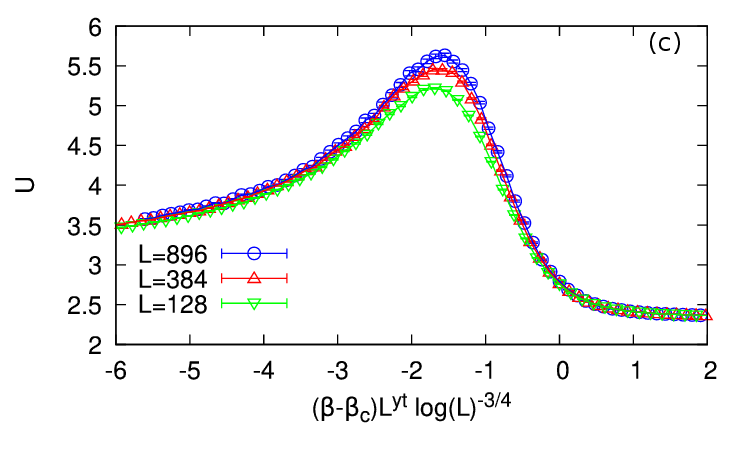}
    \caption{
        Scaling plots of Binder parameters of Potts models for (a) $q=2$, (b) $q=3$, and (c) $q=4$.
        $\beta$ is the inverse temperature and $\beta_c$ is the critical point. Exact values for critical points and exponents are used for the scaling plots.
        While the Binder parameters are monotonic functions for $q=2$,
        there are humps near the criticality on the high temperature side for $q=3$ and $4$.
    }\label{fig:binder}
\end{figure}

Consider a $q$-state ferromagnetic Potts model with $N$ spins. A spin at the site $i$ is denoted by $\sigma_i$, where $\sigma_i = 1, 2, \cdots, q$.
The Hamiltonian of this system is 
\begin{equation}
    H = - J \sum_{\left<i,j\right>} \delta_{\sigma_i, \sigma_j},
\end{equation}
where $J$ is the interaction and the summation is taken over all neighboring spins.
We define the order parameter as
\begin{equation}
    m = \frac{1}{N} \sum_i S(\sigma_i),
    \label{eq:def_m}
\end{equation}
where $S(\sigma_i)$ denotes the local quantity.
We usually adopt the following definition,
\begin{equation}
    S(\sigma_i) =\delta_{\sigma_i,1} - \frac{1}{q}.
    \label{eq:S}
\end{equation}
For the later convenience, we define the moments of the local quantity. We denote the $n$th moments of $S(\sigma_i)$ as
\begin{equation}
    I_n[S] \equiv \frac{1}{q} \sum_{\sigma_i=1}^q S(\sigma_i)^n.
\end{equation}
Note that, the local quantity $S$ is chosen so that the first moment is zero, \textit{i.e.}, $I_1[S] = 0$.

While there are several ways to define the Binder parameter, we adopt the following definition
\begin{equation}
    U(\beta, L) \equiv \frac{\left<m^4 \right>}{\left<m^2\right>^2},
\end{equation}
where $\beta$ is inverse temperature, $L$ is the system size, and $\left<\cdots \right>$ denotes the ensemble average. The finite-size scaling form of $U(\beta, L)$ is
\begin{equation}
    U(\beta, L) \sim \tilde{U}((\beta-\beta_c)L^{y_t}),
\end{equation}
where $\beta_c$ is the critical point and $y_t \equiv 1/\nu$ is the critical exponent. The order parameter $m$ can be scalar or vector as long as $m^2$ and $m^4$ are scalar values. 

Next, we introduce the improved estimator of the Binder parameter~\cite{Horita2017}. Suppose we have the Fortuin-Kasteleyn graph representation of the system~\cite{Fortuin1972}. 
The partition function of the $q$-state Potts model is written as
\begin{equation}
    Z = \sum_g v^{N_b(g)} q^{N_c(g)}, \label{eq:Z}
\end{equation}
where $v\equiv \mathrm{e}^{\beta J} -1$. $N_b(g)$ and $N_c(g)$ are the number of bonds and clusters in the graph $g$. We denote the size of the cluster of the index $k$ by $n_k$. Thus we have $N_c(g)=\sum_k 1$ and $N=\sum_k n_k$. The states of the spins in the cluster $k$ are identical and denoted by $s_k$.

The improved estimator of the 2nd moment of the order parameter is 
\begin{align}
    N^2m^2(g) & = \frac{1}{q^{N_c(g)}} \sum_{\{s_k\}} \left(\sum_k S(s_k )n_k \right)^2,                                              \\
              & \begin{aligned}
                    =\frac{1}{q^{N_c(g)}} \sum_{\{s_k\}} \Bigl( & \sum_k S(s_k)^2 n_k^2 +                           \\
                                                                & 2 \sum_{k<k'} S(s_k)S(s_{k'}) n_k n_{k'} \Bigr),
                \end{aligned}
    \\
              & = I_2[S] \sum_k n_k^2,
\end{align}
where $\sum_{\{s_k\}}$ denotes the summation over all possible spin configurations for each cluster. Here, we use the fact that the first moment is zero by definition.
The 2nd moment of the magnetization is estimated as
\begin{equation}
    N^2\left<m^2\right> = I_2[S] \left<\sum_k n_k^2 \right>_g
\end{equation}
where $\left< \cdots \right>_g$ denotes the average on graphs.
Similarly, we have the 4th moment as
\begin{align}
    N^4m^4(g) & = I_4[S] \sum_k n_k^4 + 6 I_2[S]^2 \sum_{k<k'} n_k^2 n_{k'}^2.
\end{align}
Therefore, in the graph representation, the Binder parameter is given by
\begin{equation}
    U(\beta) = A[S]
    \frac{\left< \sum_k n_k^4 \right>_g}
    {\left< \sum_k n_k^2 \right>_g^2}
    + 6 \frac{\left< \sum_{k<k'} n_k^2 n_{k'}^2 \right>_g}
    {\left< \sum_k n_k^2 \right>_g^2}, \label{eq:U42}
\end{equation}
where $A[S] \equiv I_4[S]/(I_2[S])^2$. Equation~(\ref{eq:U42}) is the improved estimator of the Binder parameter.
When we adopt Eq.~(\ref{eq:S}) for the local quantity, $A[S]$ takes
$$
    A[S] \equiv \frac{I_4[S]}{I_2[S]^2} = \frac{q^2-3q+3}{q-1} =
    \begin{cases}
        1   & q = 2, \\
        3/2 & q = 3, \\
        7/3 & q = 4. \\
    \end{cases}
$$

To investigate the behavior of the Binder parameters, we performed Monte Carlo simulations. In the following Potts model calculations, we adopted Swendseng-Wang algorithm~\cite{Swendsen1987}. After $10^4$ MCs, we observed for $10^5$ MCs. 256 independent samples are averages for each run. Throughout the manuscript, the values of the Binder parameters are estimated by the jackknife resampling method. The scaling plots for Binder parameters for various $q$ and system sizes are shown in Fig.~\ref{fig:binder}. Exact values of critical temperature and exponents are used, where $\beta_c = \log(1+\sqrt{q})$ and $y_t = 1, 6/5,$ and $3/2$ for $q=2, 3,$ and $4$. While the Binder parameter is monotonic for $q=2$, humps appear for $q>2$ and become more significant for larger values for q$q$. While the hump height seems to converge in the thermodynamic limit for $q=3$, $q=4$ exhibits the non-convergent behavior of humps.

\begin{figure}[htbp]
    \centering
    \includegraphics[width=8cm]{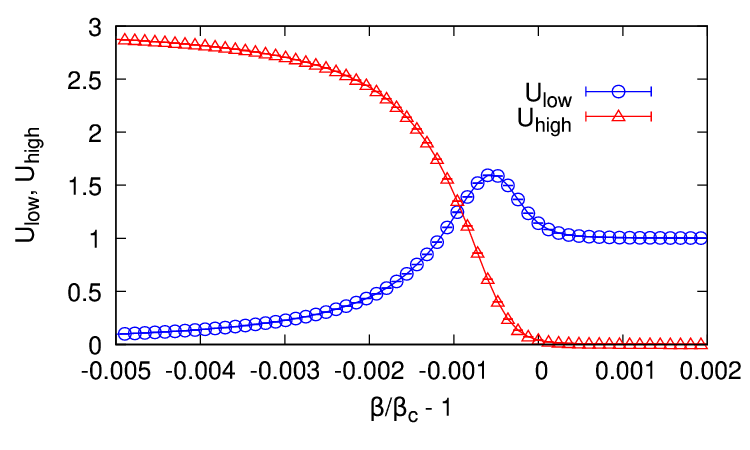}
    \caption{
        Temperature dependence of low- and high-temperature terms of the Binder parameter for $q = 3$ and $L = 896$. The exact value for the criticality point $\beta_c$ is used.
    }\label{fig:binder_component}
\end{figure}

\begin{figure}[htbp]
    \includegraphics[width=8cm]{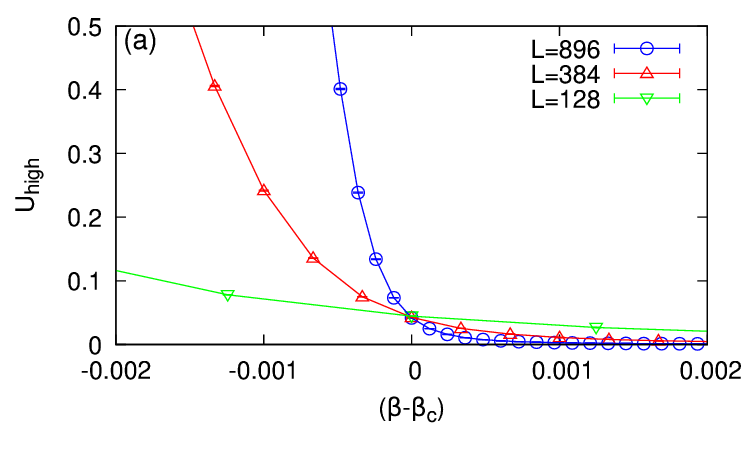}
    \includegraphics[width=8cm]{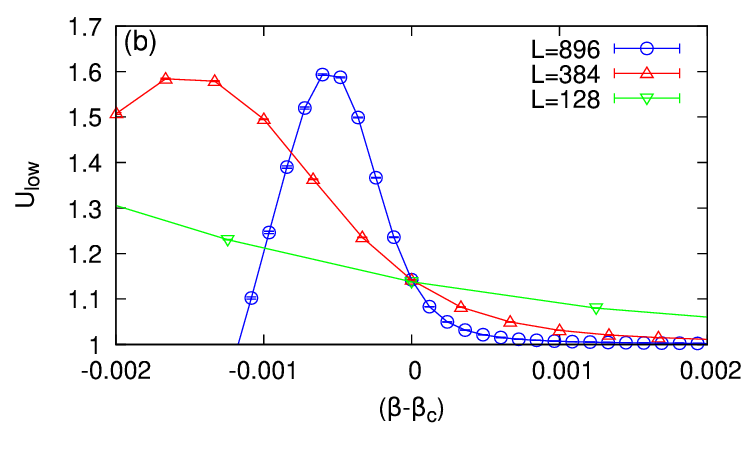}
    \caption{
        System size dependence of low- and high-temperature terms of the Binder parameter for $q=3$. (a) High-temperature terms $U_\mathrm{high}$. (b) Low-temperature terms $U_\mathrm{low}$. The exact value for the criticality point $\beta_c$ is used.
    }\label{fig:Q3_cross}
\end{figure}

As shown in Eq.~(\ref{eq:U42}), the improved estimator of the Binder parameter consists of two terms. We divide the Binder parameter into low- and high-temperature terms as 
\begin{equation}
    U = A[S] U_\mathrm{low} + U_\mathrm{high},
\end{equation}
where
\begin{equation}
    U_\mathrm{low}  \equiv  \frac{\left<\sum_k n_k^4\right>_g}{\left<\sum_k n_k^2\right>_g^2} =
    \begin{cases}
        0 & \beta \rightarrow 0,       \\
        1 & \beta \rightarrow \infty,
    \end{cases}
\end{equation}
and
\begin{equation}
    U_\mathrm{high} \equiv 6 \frac{\left<\sum_{k<k'} n_k^2 n_{k'}^2\right>_g}{\left<\sum_k n_k^2\right>_g^2} = 
    \begin{cases}
        3 & \beta \rightarrow 0,       \\
        0 & \beta \rightarrow \infty.
    \end{cases}
\end{equation}
Note that the both $U_\mathrm{low} $ and $U_\mathrm{high}$ depend only on the graph $g$, \textit{i.e.}, they do not depend on the definition of the order parameter.
The temperature dependence of each component is shown in Fig.~\ref{fig:binder_component}. One can see that the hump originates from the low-temperature term $U_\mathrm{low}$. Therefore, the non-monotonic behavior of the Binder parameter originates from the low-temperature term, which is one of the main findings of the present paper. The system size dependence of each component is shown in Fig.~\ref{fig:Q3_cross}. One can see that both components of different sizes cross at the critical point.

\begin{figure}[htbp]
    \centering
    \includegraphics[width=8cm]{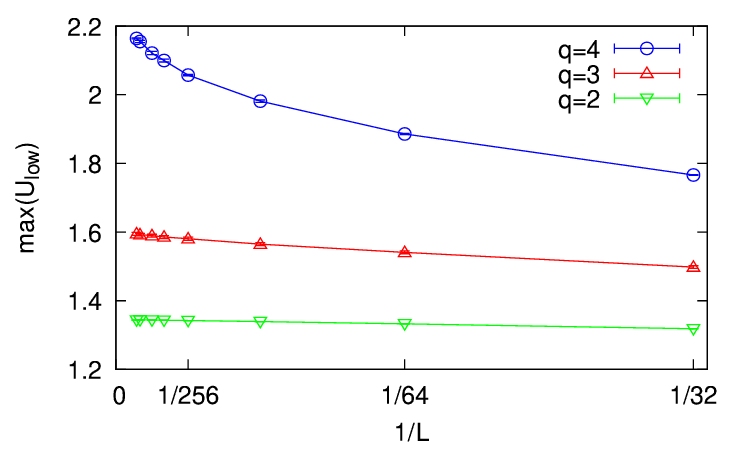}
    \caption{
        Peak height of the low-temperature term of the Binder parameter, $U_\mathrm{low}$. The cases $q=2,3$ and $4$ are shown. 
    }\label{fig:peak}
\end{figure}

The peak values of the low-temperature term increases as the system size increases. The height of the peak of $U_\mathrm{low}$ is shown in Fig.~\ref{fig:peak}. 

\begin{figure}[htbp]
    \includegraphics[width=8cm]{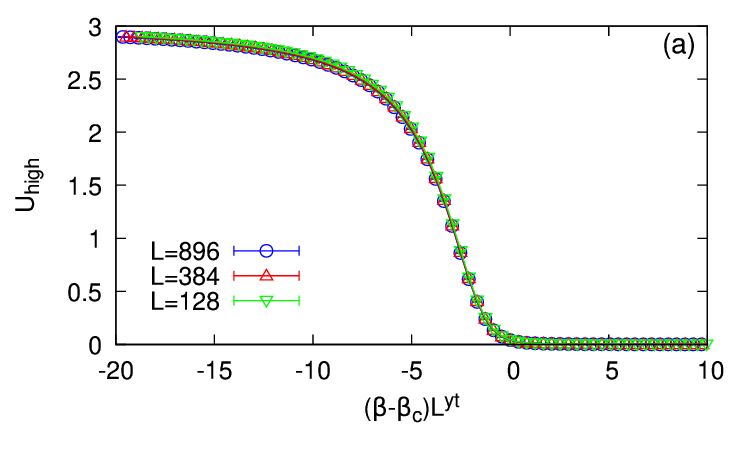}
    \includegraphics[width=8cm]{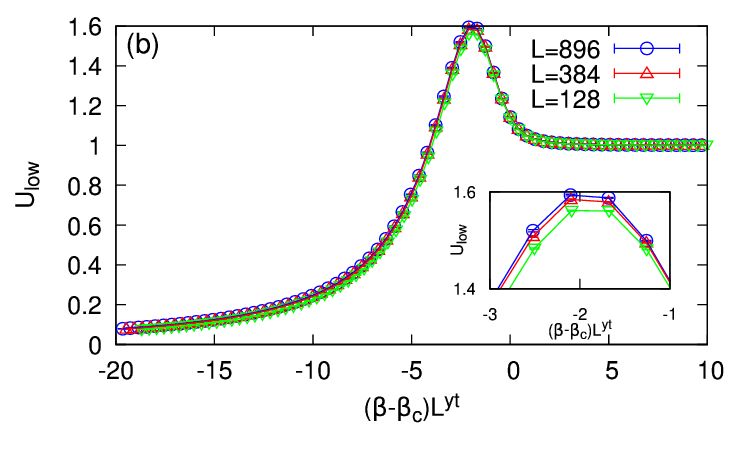}
    \caption{
        Scaling plots of (a) $U_\mathrm{high}$ and (b) $U_\mathrm{low}$ for $q=3$. Exact values for the critical point and the exponent are used for the scaling plots. The high-temperature terms $U_\mathrm{high}$ do not exhibit humps, and they are well scaled. The low-temperature terms $U_\mathrm{low}$ appear to be more size-dependent than $U_\mathrm{high}$, but it converges for larger systems. A magnified view of the area around the peak is shown in the inset.
    }\label{fig:Q3_high_low}
\end{figure}

For the case of $q=2$, the Binder parameters have no humps, but the low-temperature terms of them $U_\mathrm{low}$ have humps. However, size dependence is hardly seen. 
For the case of $q=3$, a hump appears in the Binder parameter. Furthermore, the humps become larger as the size increases but converge in the limit of large size. After the convergence, we can perform the scaling analysis for both sides of the critical point. This fact suggests that there exists the universal finite-size scaling function with a hump in $q=3$ case. Therefore, $U_\mathrm{high}$ and $U_\mathrm{low}$ are expected to be scaled simultaneously. Figure~\ref{fig:Q3_high_low} shows the scaling plots of $U_\mathrm{high}$ and $U_\mathrm{low}$ for $q=3$. 

While the height of peaks of $q=2$ and $3$ converge as the system size increases, those of $q=4$ continues to grow logarithmically, presumably due to the logarithmic correction to scaling~\cite{Fukushima2019}. Recently, Patil and Sandvik made similar plots for $q=4$ for the larger size up to $(3072 \times 3072)$, and conjectured that the peak would converge to the constant value corresponding to the Ising fluctuation~\cite{Patil2020}.

\section{Elimination of Humps} \label{sec:elimination}

Binder parameters of the discrete spin systems, such as the Potts model, sometimes exhibit humps. However, we can eliminate the humps by choosing the order parameter appropriately. For example, we can eliminate the hump for $q=4$ by adopting the Ising-like order parameter. In this section, we discusses eliminating or alleviating the humps by redefining the order parameters and Binder parameters appropriately.

\subsection{Ising order parameter}

\begin{figure}[htbp]
    \includegraphics[width=8cm]{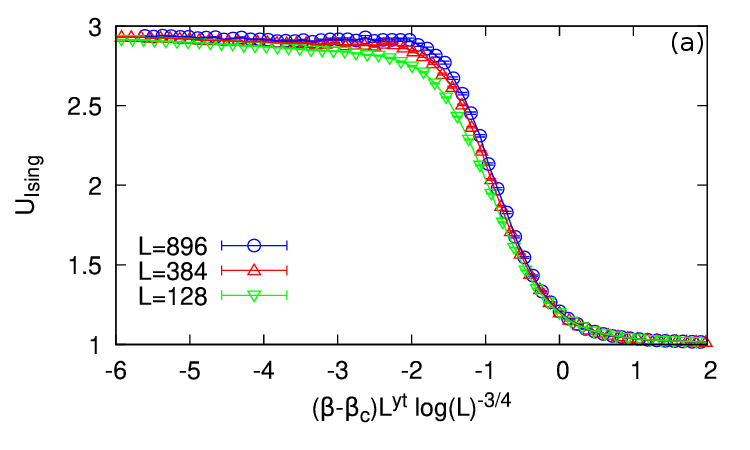}
    \includegraphics[width=8cm]{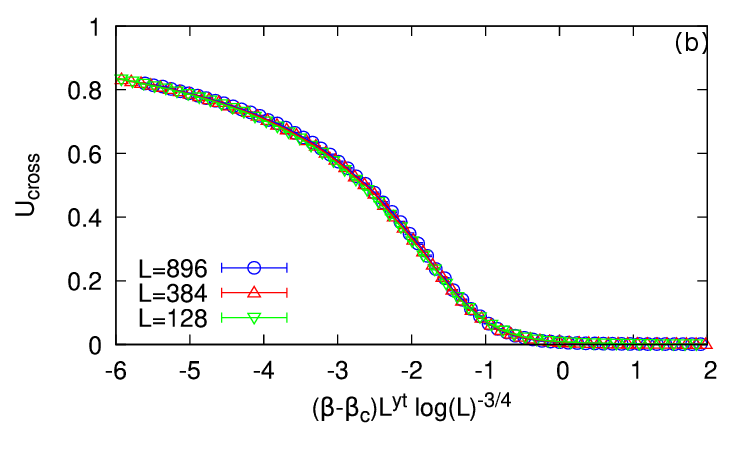}
    \caption{
        Scaling plots of Binder parameters of $q=4$ Potts model for 
        (a) Ising and (b) cross definitions. Exact values for the critical point and the exponent are used for the scaling plots.
    }\label{fig:q4_ising_cross}
\end{figure}

As shown in Fig.~\ref{fig:binder_component}, the non-monotonic behavior originates from the low-temperature term of the Binder parameter. Since the parameter $A[S]$ determines the ratio of the low-temperature term to the high-temperature term of the Binder parameter, we can reduce the effects caused by humps by reducing the value of $A[S]$. When $q$ is even, we can define an Ising-like local quantity by dividing the possible values of the spin into two groups and identifying within each group. The specific expression for the Ising-like order parameter is given by replacing Eq.~(\ref{eq:S})  with
\begin{equation}
    S_\text{Ising}(\sigma_i) = (-1)^{\sigma_i}
\end{equation}
satisfying $A[S_\text{Ising}]=1$, which is the minimum value of $A[S]$ since $I_4[S] \geq I_2[S]^2$. With the Ising order parameter, we can define the Ising Binder parameter $U_\mathrm{Ising}$. The scaling plot for the Ising Binder parameter is shown in Fig.~\ref{fig:q4_ising_cross}~(a). Although not all lines have perfectly collapsed into the single line, one can see that the scaling is better than in Fig.~\ref{fig:binder}~(c).

For $q\geq 4$, we can eliminate the low-temperature term by adopting an appropriate linear combination of two Binder parameters. Consider $q=4$ case. The improved estimators for the conventional $U$ and the Ising Binder parameter $U_\mathrm{Ising}$ are as follows.
$$
    \begin{aligned}
        U                & = \frac{7}{3} U_\mathrm{low} + U_\mathrm{high}, \\
        U_\mathrm{Ising} & = U_\mathrm{low} + U_\mathrm{high}.
    \end{aligned}
$$
Therefore, we can extract the high-temperature term of the Binder parameter $U_\mathrm{high}$ as
$$
    U_\mathrm{high} = \frac{7}{4} U_\mathrm{Ising} - \frac{3}{4} U.
$$
Since the hump originates from the low-temperature term $U_\mathrm{low}$, we can eliminate the humps by considering the high-temperature term $U_\mathrm{high}$ only. In the case of an odd number of $q$, we can define a spin variable $S$ which has a different value of $A[S]$ from the conventional definition. Then we can extract $U_\mathrm{high}$ in the similar manner. However, the above method does not work for $q=3$ since any definition of the scalar local quantity gives $A[S] = 3/2$. We will discuss it in Sec.~\ref{sec:highermoment}.

\subsection{Cross definition of Binder parameter}

For a system with a graph representation, we can express the Binder parameter in terms of cluster size moments, which gives the improved estimator. The improved estimator consists of two components, $U_\mathrm{low}$ and $U_\mathrm{high}$, and we can extract $U_\mathrm{high}$ by the linear combination of the Binder parameters calculated from several definitions of the order parameters. However, this is not possible in a system for which the improved estimator of the Binder parameter is unknown. To address this problem, we propose another Binder parameter which gives a spin representation of $U_\text{high}$.

Let us consider two kinds of local quantities, $S_\alpha(\sigma_i)$ and $S_\beta(\sigma_i)$, and define the corresponding order parameters, $m_\alpha$ and $m_\beta$, as well as Eq.~\eqref{eq:def_m}.
With the two order parameters, we define the ``cross'' definition of the Binder parameter as
\begin{equation}
    U_\mathrm{cross} \equiv \frac{\left< m_\alpha^2 m_\beta^2\right>}
    {\bigl<m_\alpha^2\bigr> \bigl<m_\beta^2\bigr>}.
    \label{eq:U_cross}
\end{equation}
In the Potts model, the graph representation of the numerator is given as
\begin{equation}
    N^4 \left(m_\alpha^2 m_\beta^2 \right)(g) =
    I_2[S_\alpha S_\beta] \sum_k n_k^4 +
    2 \left(I_2[S_\alpha] I_2[S_\beta] + 2I_1[S_\alpha S_\beta]^2\right)
    \sum_{k<k'} n_k^2 n_{k'}^2,
\end{equation}
where we assume that the first moments of $S_\alpha$ and $S_\beta$ are zero, $I_1[S_\alpha]=I_1[S_\beta]=0$.
Therefore, the cross definition of the Binder ratio is also decomposed into the low- and high-temperature terms,
\begin{equation}
    U_\text{cross} =
    \frac{I_2[S_\alpha S_\beta]}{I_2[S_\alpha]I_2[S_\beta]} U_\text{low}
    + \frac{1}{3}\left(
    1 + 2 \frac{I_1[S_\alpha S_\beta]^2}{I_2[S_\alpha]I_2[S_\beta]}
    \right) U_\text{high}.
    \label{eq:cross_decomp}
\end{equation}
In contrast to the conventional Binder ratio, the coefficient of the low-temperature term may vanish.
If two local quantities satisfy $I_2[S_\alpha S_\beta]=0$, $U_\text{cross}$ can extracts $U_\text{high}$ only.

A problem is whether such a pair of local quantities exists.
For the $q\geq 4$ Potts model, we find the simplest solution,
\begin{equation}
    S_\alpha[\sigma] \equiv \delta_{\sigma, 1} - \delta_{\sigma, 3}, \quad
    S_\beta[\sigma] \equiv \delta_{\sigma, 2} - \delta_{\sigma, 4}.
\end{equation}
Since $I_{n}[S_\alpha S_\beta] = 0$ for any positive integer $n$, we obtain $U_\text{cross}=U_\text{high}/3$.
We note that the relation, $U_\text{cross}= \left( 7 U_\text{Ising} - 3U \right) / 12$, can be proved using only the $q=4$ Potts symmetry without using the graph representation.
For large $q$, other definitions of order parameters can also satisfy the condition.
For example, we have another solution for $q=6$ as
\begin{equation*}
    S'_\alpha[\sigma] \equiv \delta_{\sigma, 1} - \delta_{\sigma, 4}, \quad
    S'_\beta[\sigma] \equiv \delta_{\sigma, 2} + \delta_{\sigma, 3}
    - \delta_{\sigma, 5} - \delta_{\sigma, 6}.
\end{equation*}
In the case of $q=3$, there is no solution of $S_\alpha$ and $S_\beta$.
It is because the coefficients of $U_\text{low}$ and $U_\text{high}$ in Eq.~\eqref{eq:cross_decomp} are always proportional.
A definition of the Binder ratio without a hump in the $q=3$ case will be discussed in the next subsection.

The scaling plot of $U_\mathrm{cross}$ in the $q=4$ Potts model on the square lattice is shown in Fig.~\ref{fig:q4_ising_cross}~(b). The cross Binder parameter is monotonic and scaled perfectly in contrast to $U$ and $U_\text{Ising}$.

The nature of $U_\text{cross}$ is unclear in a system whose graph representation is not known.
However, we expect that the cross Binder parameter defined in Eq.~\eqref{eq:U_cross} with an appropriate choice of $m_\alpha$ and $m_\beta$ exhibits monotonic behavior even in such a system.
At least the following two conditions seem to be necessary;
(i) $\left<m_\alpha^2 \right>$ and $\left<m_\beta^2 \right>$ are regarded as order parameters,
(ii) $\left<m_\alpha^2 m_\beta^2\right>$ vanishes in the low-temperature limit.
The first condition means that, in the thermodynamic limit, both of them take a non-zero value in the ordered phase while they vanish in the disordered phase.
It allows $U_\text{cross}$ to detect the phase transition.
The second condition implies that $U_\text{cross}$ goes to zero in the low temperature.
It is important so that $U_\text{cross}$ does not contain $U_\text{low}$, the cause of a hump.
To satisfy these condition, the ordered phase needs to have at least 4-fold degeneracy.
In Sec.~\ref{sec:frustratedising}, we will demonstrate that we can remove the humps by adopting $U_\text{cross}$ in a system without the graph representation. Phenomenological analysis of the first-order phase transition also supports our expectation (see Appendix \ref{sec:pheno}).

\subsection{Higher-moment Binder parameter} \label{sec:highermoment}

As shown above, a hump does not appear for $q=2$, and we can eliminate it for $q\geq 4$.
The case of $q=3$ is special. Any local quantity which satisfies $I_1[S]=0$ always leads to $A[S]=3/2$. Thus we cannot eliminate the hump by considering a linear combination of Binder parameters with different order parameters. Figure.~\ref{fig:peak} shows that the effect of the hump is expected to be small for a sufficiently large system size, but for quantum spin systems with $S_3$ symmetry, for example, the effect of the hump can be serious~\cite{Harada2013, Suzuki2015}. This subsection proposes the higher-moment Binder parameter for the three-state Potts model, which does not exhibit a hump.

\begin{figure}[htbp]
    \centering
    \includegraphics[width=8cm]{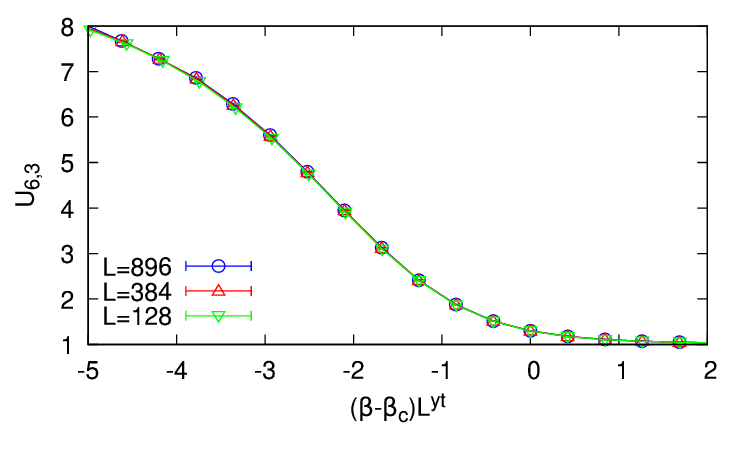}
    \caption{
        The higher-moment Binder parameters for the $q=3$ Potts model. The behavior is monotonic and exhibits better scaling plots compared with the conventional definition shown in Fig.~\ref{fig:binder}~(b). Exact values for the critical point and the exponent are used for the scaling plots.
    }\label{fig:binder63}
\end{figure}

Consider the complex order parameter for the three-state Potts model as,
$$
    S(\sigma_i) = \exp\left(i \frac{2\pi}{3} \sigma_i \right).
$$
This order parameter satisfies the following conditions for moments,
\begin{align}
    I_{3n+1}[S] = I_{3n+2}[S] & = 0,  \\
    I_{3n}[S]                 & = 1.
\end{align}
Then we define the ratio of the 6th and 3rd cumulant of the order parameters as
$$
    U_{6,3} = \frac{\left<m^6\right>}{\left<m^3\right>^2}. 
$$
The improved estimator of this parameter is given by,
\begin{equation}
    U_{6,3} = \frac{\left<\sum_k n_k^6\right>_g}{\left<\sum_k n_k^3\right>_g^2} + 20 \frac{\left< \sum_{k<k'} n_k^3 n_{k'}^3\right>_g }{\left<\sum_k n_k^3\right>_g^2}. \label{eq:binder63}
\end{equation}
The temperature dependence of the higher-moment Binder parameters are shown in Fig.~\ref{fig:binder63}. There are no any humps.

\begin{figure}[htbp]
    \centering
    \includegraphics[width=8cm]{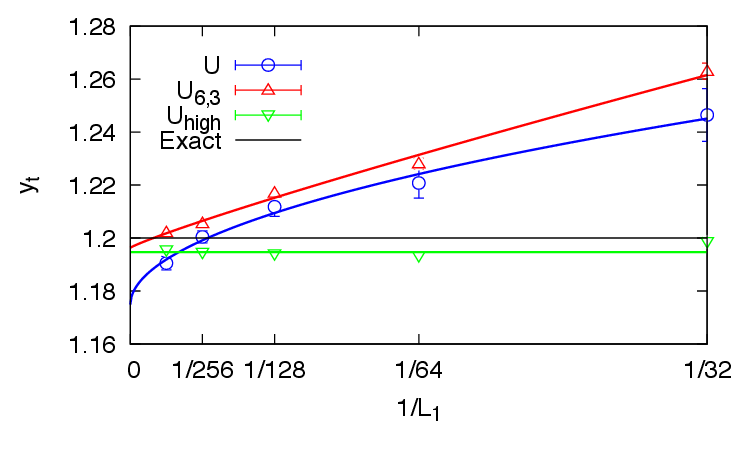}
    \caption{
        Estimation of critical exponent $y_t$ for the $q=3$ Potts model. The confidence intervals obtained by the Bayesian analysis are shown as the error bars. The curves are fitting results using Eq.~\eqref{eq:fitting}.
    }\label{fig:potts_yt}
\end{figure}

By considering the higher moment, we reduce the hump and obtain better scaling plots. To investigate how these improved scaling plots affect the estimation of the critical exponents, we perform the Bayesian scaling analysis~\cite{Harada2011,Harada2015}. 

We fixed the critical points at the exact values and estimated the critical exponents $y_t$ using a pair of two systems $(L_1, L_2)$ to demonstrate how the definition of the Binder parameter affects the finite-size scaling behavior, where $L_1 < L_2$. The results for $q=3$ Potts model is shown in Fig.~\ref{fig:potts_yt}. We used the conventional definitions $U$, the higher-moment definitions $U_{6,3}$, and the high-temperature component $U_\mathrm{high}$ to estimate the critical exponents. The pairs of system sizes are $(32, 64), (64, 128), (128, 256), (256, 512)$, and $(512, 896)$, respectively. The error bars are estimated as the confidence intervals obtained from the Bayesian scaling analysis. Reflecting the better scaling plots, the error bars of $U_{6,3}$ and $U_\mathrm{high}$ are smaller than those by $U$. Additionally, $U_\mathrm{high}$ exhibits little finite size effect. To estimate the infinite-size limit, we performed fitting by assuming the form
\begin{equation}
    y_t(L) = y_t(\infty) + a L^{-b}. \label{eq:fitting}
\end{equation}
The estimated critical exponents in infinite-size limit are summarized in Table~\ref{table:exponents} and the fitting results are shown in Fig.~\ref{fig:potts_yt}. While the critical exponent of the conventional definition deviates significantly from the exact values, the values of $U_{6,3}$ and $U_\mathrm{high}$ are in close agreement. These results suggest that the critical exponent is obtained more accurately by adopting definitions of Binder parameters without humps.

\begin{table}
    \centering
    \begin{tabular}{l|cccc}
                      & Exact & $U$     & $U_{6,3}$ & $U_\mathrm{high}$ \\
        \hline
        $y_t(\infty)$ & 1.2   & 1.17(1) & 1.196(5)  & 1.1947(5) 
    \end{tabular}
    \caption{The estimated critical exponents $y_t$ for the three-state Potts model obtained from the conventional definition $U$, the higher-moment $U_{6,3}$, and the high-temperature component $U_\mathrm{high}.$ The values of $U_{6,3}$ and $U_\mathrm{high}$ agreed with the exact value to within statistical error.}
    \label{table:exponents}
\end{table}

\section{Application to Frustrated Ising model} \label{sec:frustratedising}

In this section, we consider the frustrated $J_1$-$J_2$ model on the square lattice as a system for which the improved estimator of the Binder parameter is unknown. The Hamiltonian of this model is given by
$$
    H = J_1 \sum_{\left<i,j\right>} \sigma_i \sigma_j + J_2 \sum_{\langle \langle i,j \rangle\rangle} \sigma_i \sigma_j,
$$
where $\left<i,j\right>$ and $\langle \langle i,j \rangle\rangle$ denote the summation over the nearest and the next-nearest neighbor pairs, respectively. Spins are Ising-like variable, $\sigma_i = \pm 1$ and $J_1 <0$ and $J_2>0$ denote ferromagnetic and antiferromagnetic interactions, respectively.
The amplitude of the frustration is defined as $g = J_2/|J_1|$.
The ground state of this system is the stripe state when $g > 1/2$ and the system exhibits phase transition between the paramagnetic phase to the stripe phase at a finite temperature. While the Hamiltonian is rather simple, the nature of this transition is under debate~\cite{Sandvik2012, Kalz2012, Hong2021}. The main question is whether the system always exhibits a continuous transition in the region $g>1/2$, or there exists a point $g^*$ and the system exhibits a first-order transition in the region $1/2<g<g^*$. In this paper, we will not discuss the nature of the phase transition, but we investigate the behavior of the Binder parameter depending on the value of $g$. Since Kalz and Honecker reported that $g^*\sim 0.67$~\cite{Kalz2012}, we study two cases $g=0.55$ and $g=0.70$, respectively. When the value of $g$ is close to $0.5$, the system is expected to exhibit first-order transition properties. 

The stripe order of this system is defined as follows,
$$
    m_x \equiv \frac{1}{N} \sum_i (-1)^{x_i} \sigma_i, \quad
    m_y \equiv \frac{1}{N} \sum_i (-1)^{y_i} \sigma_i,
$$
where $(x_i, y_i)$ are the coordinates of the $i$th spin on the lattice. When the system exhibits the perfect stripe order, then the order parameter $(m_x, m_y)$ can take one of $(1,0), (-1,0), (0,1),$ and $(0,-1)$. Therefore, the ground state of this system is fourfold degenerate. There are several ways to define the Binder ratio from this order parameter $(m_x, m_y)$. The simple definition is
$$
    \begin{aligned}
        m^2 & \equiv m_x^2 + m_y^2,                                \\
        U   & \equiv \frac{\left<m^4\right>}{\left<m^2\right>^2}.
    \end{aligned}
$$
Kalz and Honecker adopted the following definition,
$$
    \begin{aligned}
        m_\mathrm{I}     & \equiv m_x + m_y,                                                          \\
        U_\mathrm{Ising} & \equiv \frac{\left<m_\mathrm{I}^4\right>}{\left<m_\mathrm{I}^2\right>^2}.
    \end{aligned}
$$
Since the scalar order parameter $m_\mathrm{I}$ can take $\pm 1$ for the perfect stripe order, this corresponds to the Ising definition which is expected to suppress the hump of the Binder parameter.
We also can define the cross definition as follows,
$$
    U_\mathrm{cross} \equiv \frac{\left<m_x^2 m_y^2\right>}{\left<m_x^2\right>\left<m_y^2\right>}.
$$
Since $\left<m_x^2 m_y^2\right>=0$ for the perfect stripe order, $U_\mathrm{cross}$ is expected to extract the high-temperature term of the Binder parameter only.

\begin{figure}[htbp]
    \includegraphics[width=8cm]{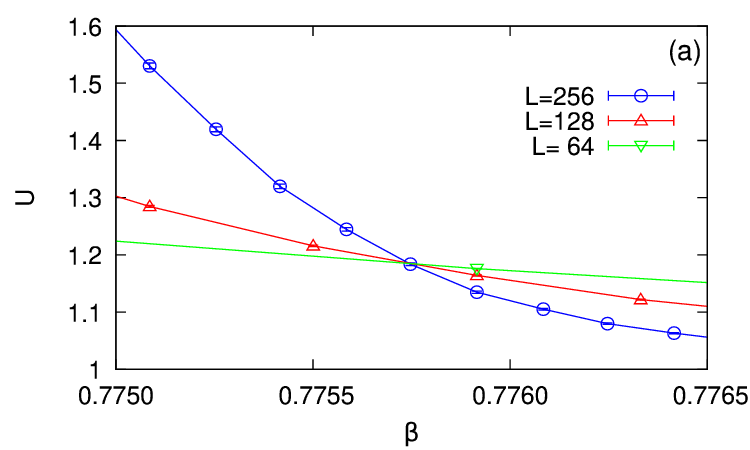}
    \includegraphics[width=8cm]{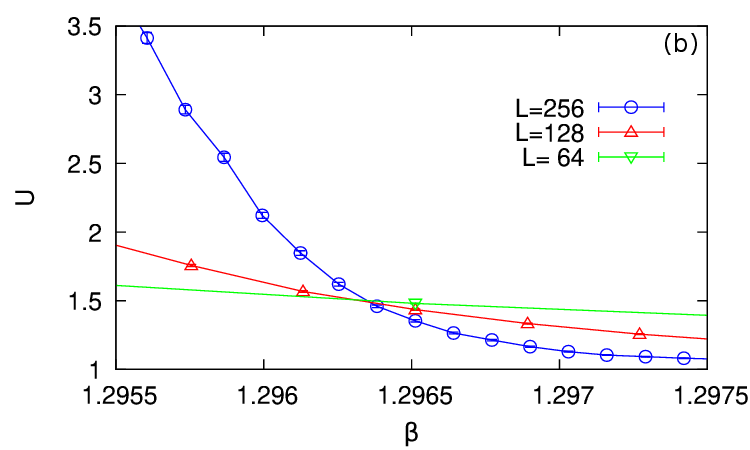}
    \caption{Size dependence of the Binder parameters with the conventional definition for
        (a) $g=0.70$ and (b) $g=0.55$. From the crossing points, we determined the critical points $\beta_c = 0.7757(1)$ for $g=0.70$ and $\beta_c = 1.2963(1)$ for $g=0.55$, respectively. 
    }\label{fig:j1j2_intersect}
\end{figure}

In the following, we performed Monte Carlo simulations with the single-spin-flip method with the Metropolis algorithm for the frustrated $J_1$-$J_2$ Ising model at two parameters, $g=0.70$ and $g=0.55$. After $10^6$ MCs, we observed for $10^6$ MCs. 256 independent samples were averaged for each run. We first determined the critical points from the crossing points of the conventional Binder parameters. The size dependence of the conventional Binder parameter is shown in Fig.~\ref{fig:j1j2_intersect}. From the obtained crossing points, we estimated the critical points $\beta_c = 0.7757(1)$ for $g=0.70$ and $\beta_c = 1.2963(1)$ for $g=0.55$. The behaviors were similar to the other definitions of the Binder parameters, and the estimated critical points were within the statistical errors.

\begin{figure}[htbp]
    \includegraphics[width=8cm]{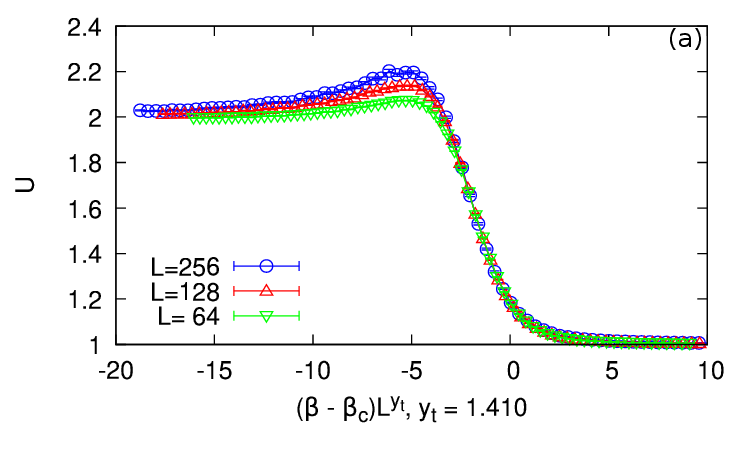}
    \includegraphics[width=8cm]{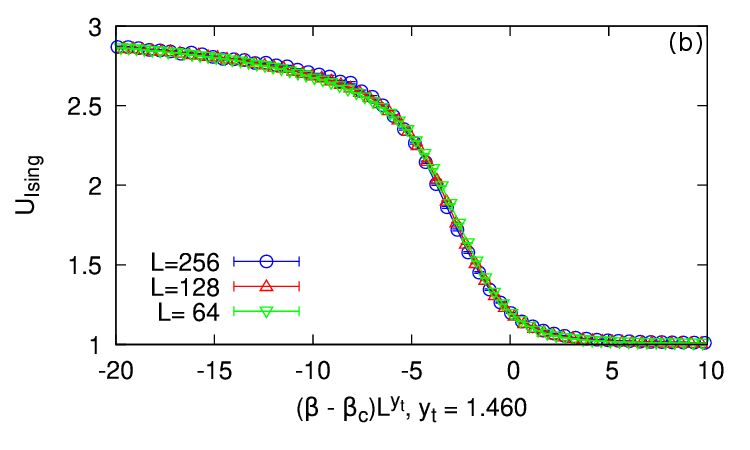}
    \includegraphics[width=8cm]{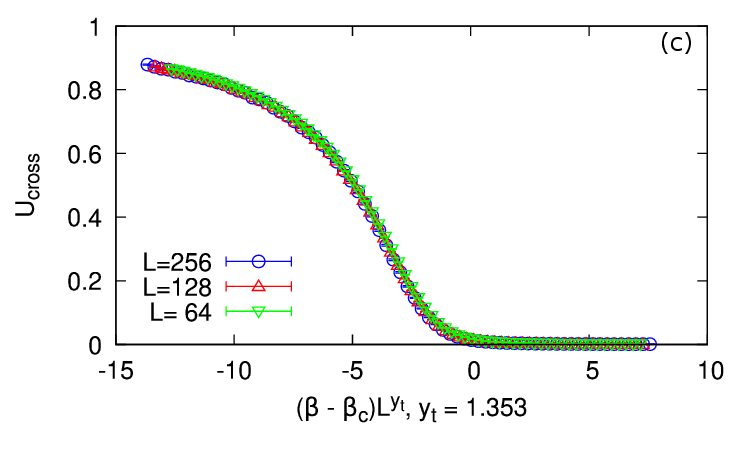}
    \caption{ Binder parameters of the frustrated $J_1$-$J_2$ Ising model for $g=0.70$
        (a) Conventional, (b) Ising, and (c) cross definitions are shown. With fixing the critical point $\beta_c = 0.7757$, we estimated the critical exponent $y_t$ for each definition by the Bayesian scaling analysis.
    }\label{fig:g070}
\end{figure}

The scaling plots of the Binder parameters for $g=0.70$ are shown in Fig.~\ref{fig:g070}. Since we can determine the critical point from the crossing point of the Binder parameter, we estimated the critical exponent $y_t$ from the Bayesian scaling analysis for each definition with fixing the critical point. While the conventional definition of the Binder parameter exhibit humps, the Ising and the cross definitions do not. As a consequence, the Binder parameters of the Ising and the cross definitions show better scaling behavior than that of the conventional one.

\begin{figure}[htbp]
    \includegraphics[width=8cm]{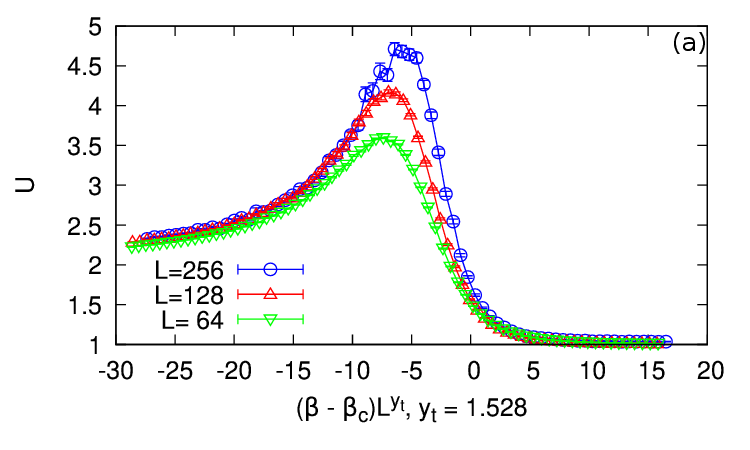}
    \includegraphics[width=8cm]{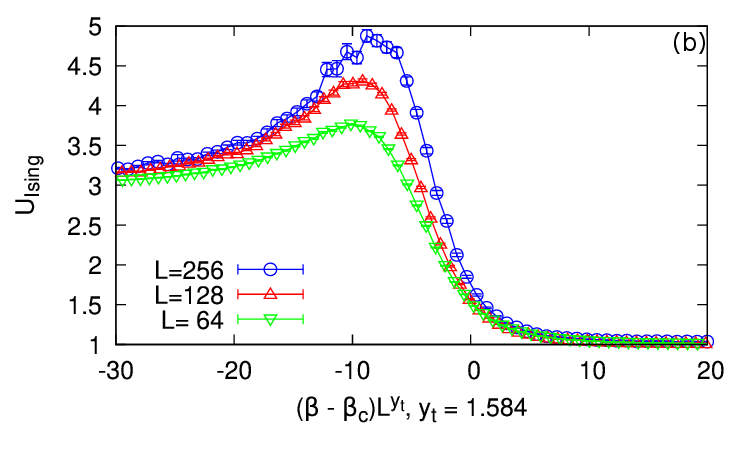}
    \includegraphics[width=8cm]{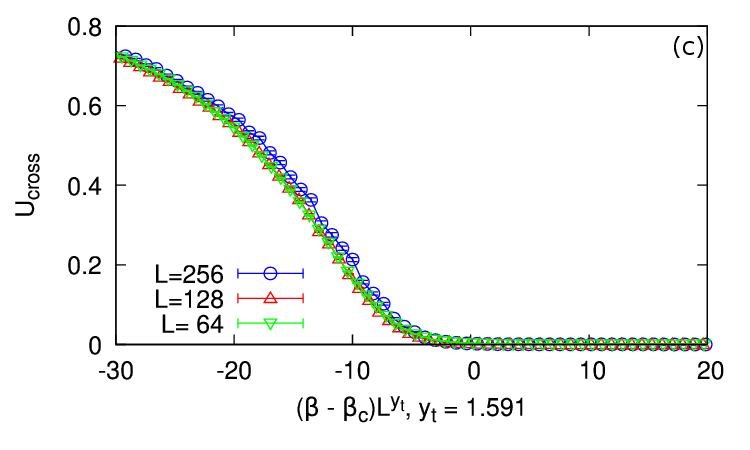}
    \caption{ Binder parameters of the frustrated $J_1$-$J_2$ Ising model for $g=0.55$
        (a) Conventional, (b) Ising, and (c) cross definitions are shown. With fixing the critical point $\beta_c = 1.2963$, we estimated the critical exponent $y_t$ for each definition by the Bayesian scaling analysis.
    }\label{fig:g055}
\end{figure}

The scaling plots of the Binder parameters for $g=0.55$ are shown in Fig.~\ref{fig:g055}. We estimated the critical exponent $y_t$ from the Bayesian scaling analysis for each definition with fixing the critical point. Unlike the case of $g=0.70$, we cannot remove humps by adopting the Ising definition. This implies that the Binder parameter \textit{feels} the first order transition in the system of $g=0.55$. The Binder parameters of the cross definition $U_\mathrm{cross}$ are monotonic and do not have any hump. So we completely removed the humps even in the systems with the first-order transitional behavior.

\begin{figure}[htbp]
    \includegraphics[width=8cm]{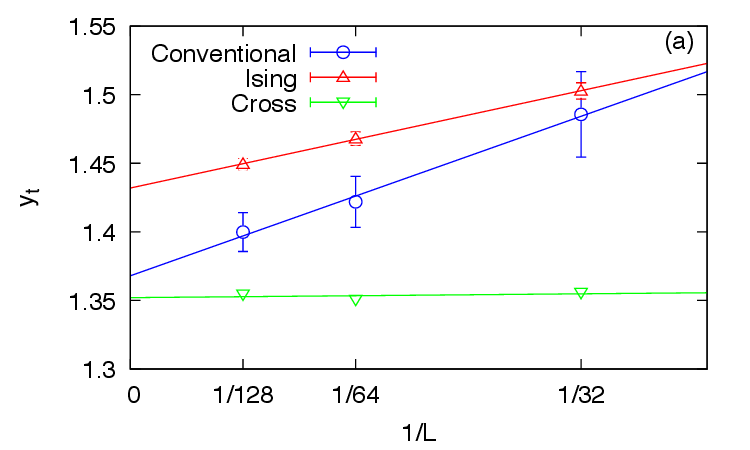}
    \includegraphics[width=8cm]{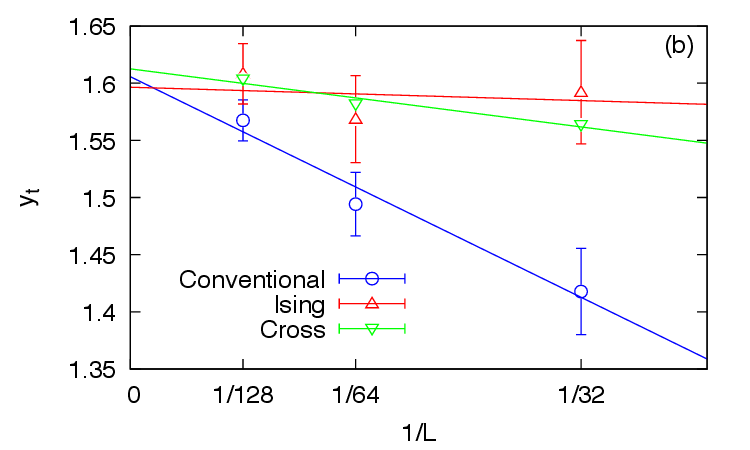}
    \caption{
        The estimated critical exponents $y_t$ of (a) $g=0.70$ and (b) $g=0.55$. The confidence intervals obtained by the Bayesian analysis are shown as the error bars.
    }\label{fig:j1j2_yt}
\end{figure}

As in the case of the Potts model, we performed the Bayesian scaling analysis of the critical exponents $y_t$ with fixing the critical points. The results are shown in Fig.~\ref{fig:j1j2_yt}. To estimate the infinite-size limit, we perform the extrapolation by assuming the form
\begin{equation}
    y_t(L) = y_t(\infty) + a L^{-1}.
\end{equation}
The extrapolation results are also shown in Fig.~\ref{fig:j1j2_yt} and the estimated values are summarized in Table~\ref{table:exponents_J1J2}. The cross definitions $U_\mathrm{cross}$ exhibit less statistical error reflecting better scaling behavior and less sensitive to the finite size effect. When $g = 0.55$, the estimates of the critical exponents by the three definitions appear to converge to the same value, whereas when $g = 070$, the value obtained by $U_\mathrm{Ising}$ deviates significantly from the values by the other two.

\begin{table}
    \centering
    \begin{tabular}{lc|ccc}
                                       &            & $U$      & $U_\mathrm{Ising}$ & $U_\mathrm{cross}$ \\
        \hline
        \multirow{2}{*}{$y_t(\infty)$} & $g = 0.70$ & 1.368(6) & 1.432(1)           & 1.352(4)           \\
                                       & $g=0.55$   & 1.61(2)  & 1.60(3)            & 1.61(1) 
    \end{tabular}
    \caption{The estimated critical exponents $y_t(\infty)$ of the $J_1-J_2$ frustrated Ising model for $g=0.70$ and $0.55$. The values obtained from the conventional definition $U$, the Ising definition $U_\mathrm{Ising}$, and the cross definition $U_\mathrm{cross}$ are shown. }
    \label{table:exponents_J1J2}
\end{table}

\section{Summary and Discussion} \label{sec:summary}

We study the hump of the Binder parameter for the discrete spin systems. The graph representation of the Binder parameter consists of two terms, high- and low-temperature terms. The hump of the Binder parameter originates from the low-temperature term. It was found that the reason why the hump is reduced by adopting the Ising definition is that the amplitude of the low-temperature term is reduced compared to the high-temperature term by changing the local definition.

We can completely eliminate the humps by extracting the high-temperature term only. We cannot find such a parameter for $q=3$. Instead, we consider the higher-moment Binder parameter, the ratio of the sixth and the third moments of the order parameter. Since the amplitude of the high-temperature term is much larger than that of the low-temperature term (see Eq.~\ref{eq:binder63}), the effect from the low-temperature term is relatively reduced. As a result, the higher-order Binder parameter does not exhibit humps. We proposed a new definition of the Binder parameter, the cross definition, which is entirely free from humps. The Binder parameter without humps exhibited better scaling plots, \textit{i.e.}, the parameters of different sizes collapsed onto a single curve for a wide temperature range. We estimated the infinite-size limit by extrapolation and found that the critical exponent obtained by the conventional definition significantly deviated from the exact value. In contrast, the values by the definitions of Binder parameters alleviating the effect of humps yielded a critical exponent with reasonable accuracy.

We applied the cross definition to the frustrated $J_1-J_2$ Ising system and found that the cross definition of the Binder parameters were monotonic. This result suggests that the humps originate from the 4th order term $\left<m^4 \right>$ of the order parameter. We performed finite-size scaling analyses, and found that the Binder parameters with different definitions also manifest finite size effects differently. Investigation of why better scaling results do not yield better estimations of critical exponents is a subject for future work.

\section*{Acknowledgements}

The computations were partially carried out on the facilities of the Supercomputer Center, Institute for Solid State Physics, University of Tokyo. We would like to thank T. Okubo, T. Suzuki, K. Harada, and S. Todo for helpful discussions. This work was supported by JSPS KAKENHI Grant Numbers JP15K05201, JP21K11923 and by MEXT as ``Exploratory Challenge on Post-K computer" (Frontiers of Basic Science: Challenging the Limits).

\appendix

\section{High-Temperature Expansion Analysis} \label{sec:HTE}

\begin{figure}[htbp]
    \centering
    \includegraphics[width=8cm]{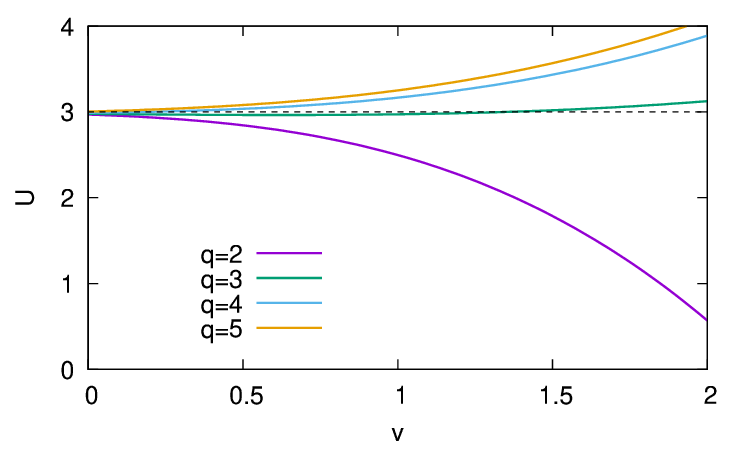}
    \caption{
        The Binder parameters with the high-temperature expansion to the third order. The cases for $q=2, 3, 4$ and $5$ are shown. The number of spins $N$ is 64. The dashed line denotes the value of the high-temperature limit.
    }\label{fig:hte}
\end{figure}

\begin{table*}[bt]
    \begin{tabular}{lllllll}
        $N_b$          & $\{n_k\}$            & $N_c$                        & $N_g$              & 
        $\sum_k n_k^2$ & $\sum_k n_k^4$       & $\sum_{k<k'} n_k^2 n_{k'}^2$                                               \\
        \hline
        $0$            & $(1,1,\dots)$        & $N$                          & $1$                & $N$ & $N$ & $N(N-1)/2$ \\
        $1$            & $(2,1,1,\dots)$      & $N-1$                        & $2N$               & 
        $N+2$          & $N+14$               & $(N-2)(N+5)/2$                                                             \\
        $2$            & $(3,1,1,\dots)$      & $N-2$                        & $6N$               & 
        $N+6$          & $N+78$               & $(N-3)(N+14)/2$                                                            \\
        $2$            & $(2,2,1,1,\dots)$    & $N-2$                        & $N(2N-7)$          & 
        $N+4$          & $N+28$               & $(N^2+7N-12)/2$                                                            \\
        $3$            & $(4,1,1,\dots)$      & $N-3$                        & $22N(=18N+4N)$     & 
        $N+12$         & $N+252$              & $(N-4)(N+27)/2$                                                            \\
        $3$            & $(3,2,1,1,\cdots)$   & $N-3$                        & $12N(N-5)$         & 
        $N+8$          & $N+92$               & $(N^2+15N-28)/2$                                                           \\
        $3$            & $(2,2,2,1,1,\cdots)$ & $N-3$                        & $2(2N^2-21N+58)/3$ & 
        $N+6$          & $N+42$               & $(N^2+11N-6)/2$

    \end{tabular}
    \caption{High-temperature expansion of the $q$-state Potts model on the square lattice. $N_b$ and $N_c$ are the number of bonds and the number of clusters in a graph. $N_g$ is the number of graphs for the cluster distribution $\{n_k\}$.}
    \label{table:3rd}
    \centering
\end{table*}

In the present manuscript, we could not show why the Binder parameter exhibits humps. However, we can rigorously show that the Binder parameter is not monotonic, and therefore, it exhibits a hump for the $Q=3$ case by the high-temperature expansion. In this Appendix, we perform the high-temperature expansion analysis of the Binder parameter in terms of $v$ in Eq.~(\ref{eq:Z}). The order of expansion corresponds to the number of bonds in the graph.

\subsection{Zeroth-order approximation}

In the zeroth order approximation, there are no bonds in the graph. The sizes of clusters are $\{n_k\} = \{1,1,\cdots,1\}$. Therefore,
\begin{align}
    N_c(g)                      & = N,                 \\
    \sum_k n_k^2 = \sum_k n_k^4 & = N,                 \\
    \sum_{k< k'} n_k^2 n_{k'}^2 & = \frac{N(N-1)}{2}.
\end{align}
The Binder parameter is
\begin{equation}
    U(v) = 3 - \frac{1}{N} (3 - A[S]) + O(v). \label{eq:zeroth}
\end{equation}
This corresponds to the high-temperature limit. For large enough system, we have
$$
    U(0) = 3.
$$

We can also have the low-temperature limit. In low-enough temperature, all spins are identical which corresponds to $N=1$ in Eq.~(\ref{eq:zeroth}). Therefore,
$$
    U(\infty) = A[S] = \frac{q^2 - 3q + 3}{q-1}.
$$
Note that the magnitude relation between the high- and low-temperature limit of the Binder parameters changes with respect to $q$ as
\begin{equation}
    \begin{cases}
        U(0) > U(\infty) & q = 2, 3, 4, \\
        U(0) < U(\infty) & q \geq 5.
    \end{cases}
\end{equation}

\subsection{First-order approximation}

There is one bond in the graph in the first order approximation. The sizes of the clusters are $\{n_k\} = \{2,1,\cdots,1\}$. Therefore,
\begin{align}
    N_c(g)                      & = N - 1,                 \\
    \sum_k n_k^2                & = N+2,                   \\
    \sum_k n_k^4                & = N+14,                  \\
    \sum_{k< k'} n_k^2 n_{k'}^2 & = \frac{(N-2)(N+5)}{2}.
\end{align}
Since the number of graphs is $2N$, the Binder parameter is
$$
    U(\beta) = 3 - \frac{1}{N} (3 - A[S]) + \frac{20}{Nq}\left(A[S] - \frac{9}{5} \right)v + O(v^2). \label{eq:first}
$$
The differential coefficient with respect to $v$ at the high-temperature limit $v = 0$ becomes positive when $A[S] > 9/5$ which corresponds to $q\geq 4$. Since $U(0) > U(\infty)$ when $q=4$, the Binder parameter must have an infection point, \textit{i.e.}, a hump. Note that we cannot prove the existence of humps for $q=3$ since the $\left. \diff U/\diff v \right|_{v=0} < 0$. We cannot show the existence of humps for $q \geq 5$ either, since the low-temperature limit of the Binder parameter is larger than the high-temperature limit.

\subsection{Third-order approximation}

We consider the high-temperature expansion to the third order on the square lattice and the results are summarized in Table~\ref{table:3rd}. The Binder parameter is
\begin{align*}
    U(v) & = 3 - \frac{3-A[S]}{N}            \\
         & + \frac{4}{Nq}(5A[S]-9) v         \\
         & +\frac{40}{N q^2} (2A[S]-3) v^2   \\
         & +\frac{16}{N q^3} (17A[S]-24) v^3 \\
         & +O\left(v^4\right).
\end{align*}
The Binder parameters expanded to the third order are shown in Fig.~\ref{fig:hte}.
While the Binder parameter of $q=3$ is decreasing function at $v = 0$, it becomes increasing function for larger $v$ while that of $q=2$ is monotonically decreasing function. These behaviors are consistent with the numerical results.

\section{Phenomenological analysis of the first-order transition} \label{sec:pheno}

In the manuscript, we showed that the cross definition of the Binder parameter $U_\mathrm{cross}$ is free from the humps, but we do not show why. In this appendix, we will show the monotonicity of $U_\mathrm{cross}$ by a phenomenological model of the first-order phase transition as well as Ref.~\cite{Vollmayr1993,Iino2019}.
We consider the two-dimensional order parameter $\bm{m}=(m_x, m_y)$ and
assume that its probability distribution consists of two parts,
\begin{equation}
    P(\bm{m}) = \frac{1}{e^{tL^d} + 1}
    \left[
        e^{tL^d} P_{>}(\bm{m}) + P_{<}(\bm{m})
        \right],
\end{equation}
where $d$, $L$, and $t$ are the space dimension, the system size, and the reduced temperature proportional to $T-T_c$.
$P_{>}(\bm{m})$ dominates for $t> 0$, and vice versa.
The order parameter in the disordered phase fluctuates around the origin.
On the other hand, $P(\bm{m})$ in the ordered phase has $q$ peaks at $\bm{m}_p = (m_{p,x}, m_{p,y})$.
Thus, the probability distributions for the disordered and ordered phase are given as
\begin{equation}
    P_{>}(\bm{m}) = \frac{1}{2\pi\sigma^2}
    \exp\left[
        - \frac{|\bm{m}|^2}{2\sigma^2}
        \right],
\end{equation}
\begin{equation}
    P_{<}(\bm{m}) = \frac{1}{2\pi\sigma^2 q}\sum_{p=1}^{q}
    \exp\left[
        - \frac{\left|\bm{m} - \bm{m}_p\right|^2}{2\sigma^2}
        \right].
\end{equation}
Here, for simplicity, we assume that the variances in $P_{<}$ and $P_{>}$ are the same.
The finite correlation length at the transition temperature implies that the variance scales with the system size as $\sigma^2=\chi_0 L^{-d}$.
The $n$th moment of the magnetization is easily calculated from the Gaussian integral.
For example, we have
\begin{equation}
    \left< m_x^2 \right>
    = \sigma^2 + \frac{1}{zq} \sum_{p=1}^q m_{p,x}^2,
\end{equation}
\begin{equation}
    \left< m_x^2 m_y^2 \right>
    = \sigma^4 + \frac{1}{zq} \sum_{p=1}^q \left(
    \sigma^2 m_{p,x}^2 + \sigma^2 m_{p,y}^2 + m_{p,x}^2 m_{p,y}^2
    \right),
\end{equation}
where $z\equiv e^{tL^d}+1$.
The odd-order moments vanish by symmetry.

\begin{figure}[htbp]
    \includegraphics[width=8cm]{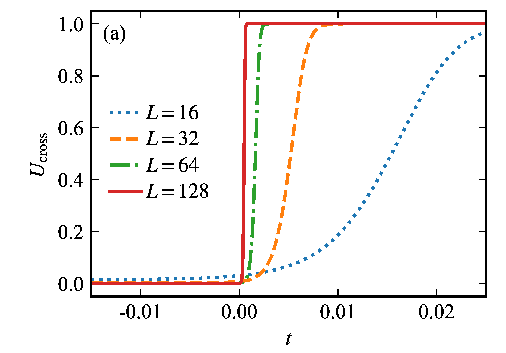}
    \includegraphics[width=8cm]{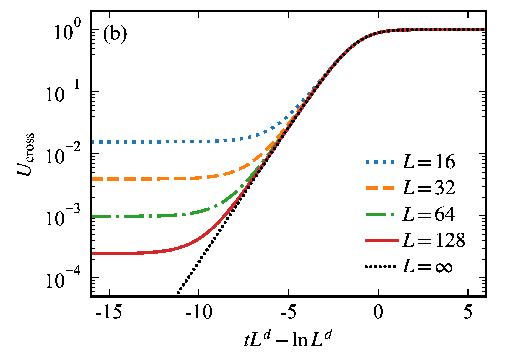}
    \caption{
        The cross definition of the Binder parameter calculated from the phenomenological model with $d=2$, $m_0=1$, and $\chi_0=1$.
        (a) $U_\text{cross}$ monotonically increases with $t$.
        (b) Scaling plot with the temperature correction collapses into the single curve.
    }\label{fig:pheno_cross}
\end{figure}

For the cross definition of the Binder parameter, it is important to choose the order parameter such that the cross term of $\bm{m}_p$ vanishes on average,
\begin{equation}
    \frac{1}{q}\sum_{p=1}^{q} m_{p,x}^2 m_{p,y}^2 = 0.
    \label{eq:m2m2}
\end{equation}
This condition erases the low-temperature part of the Binder parameter.
We consider the simplest case in which $P_{<}(\bm{m})$ has four peaks ($q=4$) at $(\pm m_0, 0)$ and $(0, \pm m_0)$.
The cross definition of the Binder parameter is calculated as
\begin{equation}
    U_\text{cross} = 1 - \left(
    \frac{m_0^2}{2z\sigma^2+m_0^2}
    \right)^2.
    \label{eq:pheno_cross}
\end{equation}
Obviously it is monotonic function of $z$.
Therefore, $U_\text{cross}$ has no hump [Fig.\ref{fig:pheno_cross}(a)].
It is not difficult to show monotonicity in more general cases which satisfy \eqref{eq:m2m2}.

The scaling plots of $U_\text{cross}$ against $tL^d$ do not collapse into a single curve.
However, by considering the temperature correction, we can put them on the same curve and obtain its scaling function.
The system-size dependence of $U_\text{cross}$ appears only in the form of $z\sigma^2$.
Let us find the temperature $t_0$ at which $z\sigma^2$ is equal to a constant value.
The leading order of $t_0$ satisfies $t_0 L^d \simeq \ln L^d + O(1)$.
Thus, if the abscissa of the scaling plot is $tL^d-\ln L^d$, all the curves intersect at a single point.
This shift corresponds to a correction of the transition temperature at each size,
\begin{equation}
    T_c(L) = T_c(\infty) \left(1 + L^{-d}\ln L^d\right).
\end{equation}
It implies that the finite-scaling analysis of $U_\text{cross}$ may overestimate the transition temperature.
We emphasize that this correction is different from and much smaller than the logarithmic correction in the 4-state Potts model~\cite{Salas1997}.

As shown in Fig.\ref{fig:pheno_cross}(b), the scaling plots of Eq.\eqref{eq:pheno_cross} with the temperature correction are on the same curve.
Its scaling function in the thermodynamic limit is obtained as
\begin{equation}
    U_\text{cross} \simeq 1 - \left(
    \frac{m_0^2}{2\chi_0 e^x + m_0^2}
    \right)^2
\end{equation}
with $x\equiv tL^d-\ln L^d$.
We note that $U_\text{cross}$ is scaled well in the entire region above a certain temperature in contrast to the conventional and Ising definitions.
These exhibit data collapse only near the transition temperature because a hump exists at the high temperature, as well as the clock-type definition of the Binder parameter~\cite{Iino2019}.

\bibliography{binder_potts}

\begin{thebibliography}{21}
\expandafter\ifx\csname natexlab\endcsname\relax\def\natexlab#1{#1}\fi
\expandafter\ifx\csname bibnamefont\endcsname\relax
  \def\bibnamefont#1{#1}\fi
\expandafter\ifx\csname bibfnamefont\endcsname\relax
  \def\bibfnamefont#1{#1}\fi
\expandafter\ifx\csname citenamefont\endcsname\relax
  \def\citenamefont#1{#1}\fi
\expandafter\ifx\csname url\endcsname\relax
  \def\url#1{\texttt{#1}}\fi
\expandafter\ifx\csname urlprefix\endcsname\relax\def\urlprefix{URL }\fi
\providecommand{\bibinfo}[2]{#2}
\providecommand{\eprint}[2][]{\url{#2}}

\bibitem[{\citenamefont{Binder}(1981)}]{Binder1981}
\bibinfo{author}{\bibfnamefont{K.}~\bibnamefont{Binder}}, \bibinfo{journal}{Z.
  Phys., B Condens. matter} \textbf{\bibinfo{volume}{43}}, \bibinfo{pages}{119}
  (\bibinfo{year}{1981}).

\bibitem[{\citenamefont{Binder et~al.}(1985)\citenamefont{Binder, Nauenberg,
  Privman, and Young}}]{Binder1985}
\bibinfo{author}{\bibfnamefont{K.}~\bibnamefont{Binder}},
  \bibinfo{author}{\bibfnamefont{M.}~\bibnamefont{Nauenberg}},
  \bibinfo{author}{\bibfnamefont{V.}~\bibnamefont{Privman}}, \bibnamefont{and}
  \bibinfo{author}{\bibfnamefont{A.~P.} \bibnamefont{Young}},
  \bibinfo{journal}{Phys. Rev. B} \textbf{\bibinfo{volume}{31}},
  \bibinfo{pages}{1498} (\bibinfo{year}{1985}).

\bibitem[{\citenamefont{Watanabe et~al.}(2012)\citenamefont{Watanabe, Ito, and
  Hu}}]{Watanabe2012}
\bibinfo{author}{\bibfnamefont{H.}~\bibnamefont{Watanabe}},
  \bibinfo{author}{\bibfnamefont{N.}~\bibnamefont{Ito}}, \bibnamefont{and}
  \bibinfo{author}{\bibfnamefont{C.-K.} \bibnamefont{Hu}}, \bibinfo{journal}{J.
  Chem. Phys.} \textbf{\bibinfo{volume}{136}}, \bibinfo{pages}{204102}
  (\bibinfo{year}{2012}).

\bibitem[{\citenamefont{Morita and Kawashima}(2019)}]{Morita2019}
\bibinfo{author}{\bibfnamefont{S.}~\bibnamefont{Morita}} \bibnamefont{and}
  \bibinfo{author}{\bibfnamefont{N.}~\bibnamefont{Kawashima}},
  \bibinfo{journal}{Comput. Phys. Commun.} \textbf{\bibinfo{volume}{236}},
  \bibinfo{pages}{65} (\bibinfo{year}{2019}).

\bibitem[{\citenamefont{Hasenbusch}(2008)}]{Hasenbusch2008}
\bibinfo{author}{\bibfnamefont{M.}~\bibnamefont{Hasenbusch}},
  \bibinfo{journal}{J. Stat. Mech. Theory Exp.}
  \textbf{\bibinfo{volume}{2008}}, \bibinfo{pages}{P08003}
  (\bibinfo{year}{2008}).

\bibitem[{\citenamefont{Tomita and Okabe}(2002)}]{Tomita2002}
\bibinfo{author}{\bibfnamefont{Y.}~\bibnamefont{Tomita}} \bibnamefont{and}
  \bibinfo{author}{\bibfnamefont{Y.}~\bibnamefont{Okabe}},
  \bibinfo{journal}{Phys. Rev. B} \textbf{\bibinfo{volume}{66}},
  \bibinfo{pages}{180401} (\bibinfo{year}{2002}).

\bibitem[{\citenamefont{Jin et~al.}(2012)\citenamefont{Jin, Sen, and
  Sandvik}}]{Sandvik2012}
\bibinfo{author}{\bibfnamefont{S.}~\bibnamefont{Jin}},
  \bibinfo{author}{\bibfnamefont{A.}~\bibnamefont{Sen}}, \bibnamefont{and}
  \bibinfo{author}{\bibfnamefont{A.~W.} \bibnamefont{Sandvik}},
  \bibinfo{journal}{Phys. Rev. Lett.} \textbf{\bibinfo{volume}{108}},
  \bibinfo{pages}{045702} (\bibinfo{year}{2012}).

\bibitem[{\citenamefont{Kalz and Honecker}(2012)}]{Kalz2012}
\bibinfo{author}{\bibfnamefont{A.}~\bibnamefont{Kalz}} \bibnamefont{and}
  \bibinfo{author}{\bibfnamefont{A.}~\bibnamefont{Honecker}},
  \bibinfo{journal}{Phys. Rev. B} \textbf{\bibinfo{volume}{86}},
  \bibinfo{pages}{134410} (\bibinfo{year}{2012}).

\bibitem[{\citenamefont{Harada et~al.}(2013)\citenamefont{Harada, Suzuki,
  Okubo, Matsuo, Lou, Watanabe, Todo, and Kawashima}}]{Harada2013}
\bibinfo{author}{\bibfnamefont{K.}~\bibnamefont{Harada}},
  \bibinfo{author}{\bibfnamefont{T.}~\bibnamefont{Suzuki}},
  \bibinfo{author}{\bibfnamefont{T.}~\bibnamefont{Okubo}},
  \bibinfo{author}{\bibfnamefont{H.}~\bibnamefont{Matsuo}},
  \bibinfo{author}{\bibfnamefont{J.}~\bibnamefont{Lou}},
  \bibinfo{author}{\bibfnamefont{H.}~\bibnamefont{Watanabe}},
  \bibinfo{author}{\bibfnamefont{S.}~\bibnamefont{Todo}}, \bibnamefont{and}
  \bibinfo{author}{\bibfnamefont{N.}~\bibnamefont{Kawashima}},
  \bibinfo{journal}{Phys. Rev. B} \textbf{\bibinfo{volume}{88}},
  \bibinfo{pages}{220408} (\bibinfo{year}{2013}).

\bibitem[{\citenamefont{Suzuki et~al.}(2015)\citenamefont{Suzuki, Harada,
  Matsuo, Todo, and Kawashima}}]{Suzuki2015}
\bibinfo{author}{\bibfnamefont{T.}~\bibnamefont{Suzuki}},
  \bibinfo{author}{\bibfnamefont{K.}~\bibnamefont{Harada}},
  \bibinfo{author}{\bibfnamefont{H.}~\bibnamefont{Matsuo}},
  \bibinfo{author}{\bibfnamefont{S.}~\bibnamefont{Todo}}, \bibnamefont{and}
  \bibinfo{author}{\bibfnamefont{N.}~\bibnamefont{Kawashima}},
  \bibinfo{journal}{Phys. Rev. B} \textbf{\bibinfo{volume}{91}},
  \bibinfo{pages}{094414} (\bibinfo{year}{2015}).

\bibitem[{\citenamefont{Vollmayr et~al.}(1993)\citenamefont{Vollmayr, Reger,
  Scheucher, and Binder}}]{Vollmayr1993}
\bibinfo{author}{\bibfnamefont{K.}~\bibnamefont{Vollmayr}},
  \bibinfo{author}{\bibfnamefont{J.~D.} \bibnamefont{Reger}},
  \bibinfo{author}{\bibfnamefont{M.}~\bibnamefont{Scheucher}},
  \bibnamefont{and} \bibinfo{author}{\bibfnamefont{K.}~\bibnamefont{Binder}},
  \bibinfo{journal}{Z. Phys., B Condens. matter} \textbf{\bibinfo{volume}{91}},
  \bibinfo{pages}{113} (\bibinfo{year}{1993}).

\bibitem[{\citenamefont{Patil and Sandvik}(2020)}]{Patil2020}
\bibinfo{author}{\bibfnamefont{P.}~\bibnamefont{Patil}} \bibnamefont{and}
  \bibinfo{author}{\bibfnamefont{A.~W.} \bibnamefont{Sandvik}},
  \bibinfo{journal}{Phys. Rev. B} \textbf{\bibinfo{volume}{101}},
  \bibinfo{pages}{014453} (\bibinfo{year}{2020}).

\bibitem[{\citenamefont{Horita et~al.}(2017)\citenamefont{Horita, Suwa, and
  Todo}}]{Horita2017}
\bibinfo{author}{\bibfnamefont{T.}~\bibnamefont{Horita}},
  \bibinfo{author}{\bibfnamefont{H.}~\bibnamefont{Suwa}}, \bibnamefont{and}
  \bibinfo{author}{\bibfnamefont{S.}~\bibnamefont{Todo}},
  \bibinfo{journal}{Phys. Rev. E} \textbf{\bibinfo{volume}{95}},
  \bibinfo{pages}{012143} (\bibinfo{year}{2017}).

\bibitem[{\citenamefont{Fortuin and Kasteleyn}(1972)}]{Fortuin1972}
\bibinfo{author}{\bibfnamefont{C.}~\bibnamefont{Fortuin}} \bibnamefont{and}
  \bibinfo{author}{\bibfnamefont{P.}~\bibnamefont{Kasteleyn}},
  \bibinfo{journal}{Physica} \textbf{\bibinfo{volume}{57}}, \bibinfo{pages}{536
  } (\bibinfo{year}{1972}).

\bibitem[{\citenamefont{Swendsen and Wang}(1987)}]{Swendsen1987}
\bibinfo{author}{\bibfnamefont{R.~H.} \bibnamefont{Swendsen}} \bibnamefont{and}
  \bibinfo{author}{\bibfnamefont{J.-S.} \bibnamefont{Wang}},
  \bibinfo{journal}{Phys. Rev. Lett.} \textbf{\bibinfo{volume}{58}},
  \bibinfo{pages}{86} (\bibinfo{year}{1987}).

\bibitem[{\citenamefont{Fukushima and Sakai}(2019)}]{Fukushima2019}
\bibinfo{author}{\bibfnamefont{K.}~\bibnamefont{Fukushima}} \bibnamefont{and}
  \bibinfo{author}{\bibfnamefont{K.}~\bibnamefont{Sakai}},
  \bibinfo{journal}{Prog. Theor. Exp. Phys.} \textbf{\bibinfo{volume}{2019}}
  (\bibinfo{year}{2019}).

\bibitem[{\citenamefont{Harada}(2011)}]{Harada2011}
\bibinfo{author}{\bibfnamefont{K.}~\bibnamefont{Harada}},
  \bibinfo{journal}{Phys. Rev. E} \textbf{\bibinfo{volume}{84}},
  \bibinfo{pages}{056704} (\bibinfo{year}{2011}).

\bibitem[{\citenamefont{Harada}(2015)}]{Harada2015}
\bibinfo{author}{\bibfnamefont{K.}~\bibnamefont{Harada}},
  \bibinfo{journal}{Phys. Rev. E} \textbf{\bibinfo{volume}{92}},
  \bibinfo{pages}{012106} (\bibinfo{year}{2015}).

\bibitem[{\citenamefont{Li and Yang}(2021)}]{Hong2021}
\bibinfo{author}{\bibfnamefont{H.}~\bibnamefont{Li}} \bibnamefont{and}
  \bibinfo{author}{\bibfnamefont{L.-P.} \bibnamefont{Yang}},
  \bibinfo{journal}{Phys. Rev. E} \textbf{\bibinfo{volume}{104}},
  \bibinfo{pages}{024118} (\bibinfo{year}{2021}).

\bibitem[{\citenamefont{Iino et~al.}(2019)\citenamefont{Iino, Morita,
  Kawashima, and Sandvik}}]{Iino2019}
\bibinfo{author}{\bibfnamefont{S.}~\bibnamefont{Iino}},
  \bibinfo{author}{\bibfnamefont{S.}~\bibnamefont{Morita}},
  \bibinfo{author}{\bibfnamefont{N.}~\bibnamefont{Kawashima}},
  \bibnamefont{and} \bibinfo{author}{\bibfnamefont{A.~W.}
  \bibnamefont{Sandvik}}, \bibinfo{journal}{J. Phys. Soc. Jpn}
  \textbf{\bibinfo{volume}{88}}, \bibinfo{pages}{034006}
  (\bibinfo{year}{2019}).

\bibitem[{\citenamefont{Salas and Sokal}(1997)}]{Salas1997}
\bibinfo{author}{\bibfnamefont{J.}~\bibnamefont{Salas}} \bibnamefont{and}
  \bibinfo{author}{\bibfnamefont{A.~D.} \bibnamefont{Sokal}},
  \bibinfo{journal}{J. Stat. Phys.} \textbf{\bibinfo{volume}{88}},
  \bibinfo{pages}{567} (\bibinfo{year}{1997}), \eprint{9607030}.

\end{thebibliography}

\end{document}